\documentstyle[12pt]{article}
\begin{document}
\input FEYNMAN
\newcommand{\beq}{\begin{equation}}
\newcommand{\eeq}{\end{equation}}
\newcommand{\bea}{\begin{eqnarray}}
\newcommand{\eea}{\end{eqnarray}}
\newcommand{\dirac}{/\!\!\!\partial}
\newcommand{\Dirac}{/\!\!\!\!D}
\def\lequiv{\raise 0.4ex \hbox{$<$} \kern -0.8 em \lower 0.62 ex \hbox{$\sim$}
}
\def\gequiv{\raise 0.4ex \hbox{$>$} \kern -0.7 em \lower 0.62 ex \hbox{$\sim$}
}
\renewcommand{\theequation}{\thesection.\arabic{equation}}
\begin{titlepage}
\begin{center}
\hfill hep-th/9408073~~~~ IASSNS-HEP-94/63\\ \hfill NYU-TH-94/05/01~~~~
CERN-TH.7388/94
\vskip .2in
{\large\bf Tree-Level Unitarity Constraints on the Gravitational Couplings of
Higher-Spin Massive Fields}
\end{center}
\vskip .2in
\begin{center}
{\large A. Cucchieri,~~~ M. Porrati\footnotemark}
\vskip .1in
{\em Department of Physics, NYU,
4 Washington Pl., New York, NY 10003, USA}
\footnotetext{Also at: Theory Division,
CERN, CH-1211 Geneva 23, Switzerland}
\vskip .1in and
\vskip .1in
{\large S. Deser}\footnotemark
\footnotetext{Permanent address:
Department of Physics, Brandeis Univ. Waltham, MA 02254-9110,
USA.}
\vskip .1in
{\em Institute for Advanced Study, School of Natural Sciences,
Princeton, NJ 08540, USA\\ and \\
Theory Division, CERN, CH-1211 Geneva 23, Switzerland}
\vskip .3in
{\bf ABSTRACT} \end{center}
\begin{quotation}
\noindent
We analyse the high-energy behavior of tree-level graviton Compton amplitudes
for particles of mass $m$ and arbitrary spin, concentrating on a combination of
forward amplitudes that will be unaffected
by eventual cross-couplings to other, higher spins.
We first show that for any spin $>$ 2, tree-level unitarity is already violated
at energies $\sim$ $\sqrt{mM_{Pl}}$, rather than at the Planck scale $M_{Pl}$,
even for $ m \ll M_{Pl}$. We then restore unitarity to this amplitude
up to $M_{Pl}$ by adding
non-minimal couplings that depend on the curvature and its derivatives, and
modify the minimal description -- including particle
gravitational quadrupole moments -- at
$\sim m^{-1}$ scales.
\end{quotation}
\vskip .1in
\begin{center} {\em Dedicated to the Memory of Julian Schwinger} \end{center}
\vfill
CERN-TH.7388/94 \\ August, 1994
\end{titlepage}
\section{Introduction}
It is fortunate that no higher ($s>2$) spin elementary excitations have been
observed: massless ones are well-known to suffer from grave consistency
problems when they -- unavoidably -- couple
to gravity\footnotemark~\cite{AD1};
\footnotetext{Consistency may be achieved in the presence of a negative
cosmological constant $\Lambda$~\cite{Vass}. Here we deal only with $
\Lambda=0$ gravity.}
massive ones,
with which we will be concerned here, are not directly touched by this
difficulty, related to the loss of free-field gauge invariance when ordinary
derivatives are replaced by covariant ones. Instead, we shall see that they
inherit from their $m=0$ parts an obstacle of a more intuitive kind, namely
that tree-level unitarity is already violated at energies $\sqrt{\mbox{s}} \sim
\sqrt{mM_{Pl}}$, well below Planck mass $M_{Pl}$ for ordinary $m$ in e.g.
graviton Compton scattering. One of the motivations for this study is the fact
that massive excitations of any spin are necessarily present in the context of
string theory, at least at levels where a particle expansion is meaningful.
Those masses can be smaller than the Planck scale if the string is weakly
interacting; thus, one would like to see how field theory arranges itself
``optimally'' in this context and compare this with the description arising
from the beta-function expansion, say, of the string.

A very similar problem arises in the electromagnetic coupling of higher-spin
charged fields, and provides another motivation for the gravitational analysis.
In the Maxwell case, it was found~\cite{FPT} that judicious addition of
non-minimal, Pauli, couplings restores low-energy tree unitarity in Compton
scattering with the added bonus that the new terms give rise to a universal
value $g=2$ of the $g$-factor for all spins. The corresponding question here
would be about the gravitational quadrupole moments of the matter system.
The first example, $s=5/2$, of our topic was treated recently~\cite{P}; both
the gravitational Compton amplitude difficulty and its (partial) cure were
exhibited. Here we shall generalize these results to cover all bosonic and
fermionic spins $\geq 5/2$.
We shall also emphasize that there are in fact two very different contributions
to the graviton-particle vertex or current, defined with respect to the gauge
transformations permitted by the massless parts. Here we shall deal with the
``gauge-longitudinal'' parts of the currents, where our non-minimal completion
will indeed improve the high-energy behavior. These new terms will also
contribute to (but not remove) the unwanted behavior of the gauge-transverse
currents;
we conjecture that more radical means will be required here, including (at
least) addition of infinite Regge towers of higher-spin modes, and
corresponding
corrections to the Einstein action, as automatically happens in string theory.

Let us briefly review the problems of and differences
between massive and massless spin$>1/2$ fields.
The field equations of the latter are
always of the schematic form $D_\mu F^{\mu\nu...}=0$ for bosons, where $F$ is
the antisymmetric field strength. Hence the Bianchi identity $2D_\nu D_\mu
F^{\mu\nu...}=[D_\nu,D_\mu]F^{\mu\nu...}$ will lead to unacceptable local
constraints of the form $RF=0$ on the matter field and full
curvature tensor $R$
(only spin 1 and 2 escape this difficulty).
A similar problem occurs for fermions,
where $F$ is proportional to the fermionic potential itself (here only spin 3/2
escapes because just the Ricci tensor enters in the above constraint).
In the massless case, there can be no non-minimal help for these problems, by
gauge invariance and dimensions.
For massive models, on the other hand, the loss of Bianchi
identities is not as catastrophic because the resulting constraints simply
relate the $RF$ term to terms $\sim m^2 D_\mu \phi^{\mu...}$, where $\phi$ is
the bosonic amplitude, for example, thereby merely shifting the usual
$m^2\partial_\mu \phi^{\mu...}=0$ condition of flat space. Furthermore, once a
mass parameter is present, it becomes possible to add terms such as
$\sim m^{-1} R
\bar{\psi}\psi$ (for fermions), a mechanism we will indeed exploit here.
There are two equivalent ways of performing the analysis: either directly, at
the level of the matter field equations (where one can in turn either attempt
an all-order analysis or expand in powers of the metric deviation
$h_{\mu\nu}=g_{\mu\nu}-\eta_{\mu\nu}$ from flatness), or by considering the
Compton scattering amplitude, where the field equations' characteristics are
divided into their linear parts in $h_{\mu\nu}$ (the vertices) and their
flat-space parts (the propagators). Here we proceed in the latter way, the
fundamental difference, for our purposes, between $s\leq 2$ and
$s>2$ fields being that, for the former, the scattering amplitudes
remain small up to the Planck scale.
In order to convince ourselves that this lower-spin property
is not obvious, let us consider, for instance, the elastic
scattering of a graviton off a massive
spin-$s$ particle, depicted in fig.~1.
\\
\begin{picture}(100000,16500)
\drawline\fermion[\SE\REG](0,15500)[5000]
\drawline\scalar[\NE\REG](\particlebackx,\particlebacky)[3]
\drawline\fermion[\S\REG](\scalarfrontx,\scalarfronty)[7500]
\drawline\scalar[\SE\REG](\particlebackx,\particlebacky)[3]
\drawline\fermion[\SW\REG](\scalarfrontx,\scalarfronty)[5000]
\drawline\fermion[\SE\REG](19000,15500)[5000]
\drawline\scalar[\SE\REG](\particlebackx,\particlebacky)[6]
\drawline\fermion[\S\REG](\scalarfrontx,\scalarfronty)[7500]
\drawline\scalar[\NE\REG](\particlebackx,\particlebacky)[6]
\drawline\fermion[\SW\REG](\scalarfrontx,\scalarfronty)[5000]
\drawline\fermion[\NW\REG](41500,8550)[6000]
\drawline\scalar[\NE\REG](\pfrontx,\pfronty)[4]
\drawline\scalar[\SE\REG](\particlefrontx,\particlefronty)[4]
\drawline\fermion[\SW\REG](\particlefrontx,\particlefronty)[6000]
\end{picture}
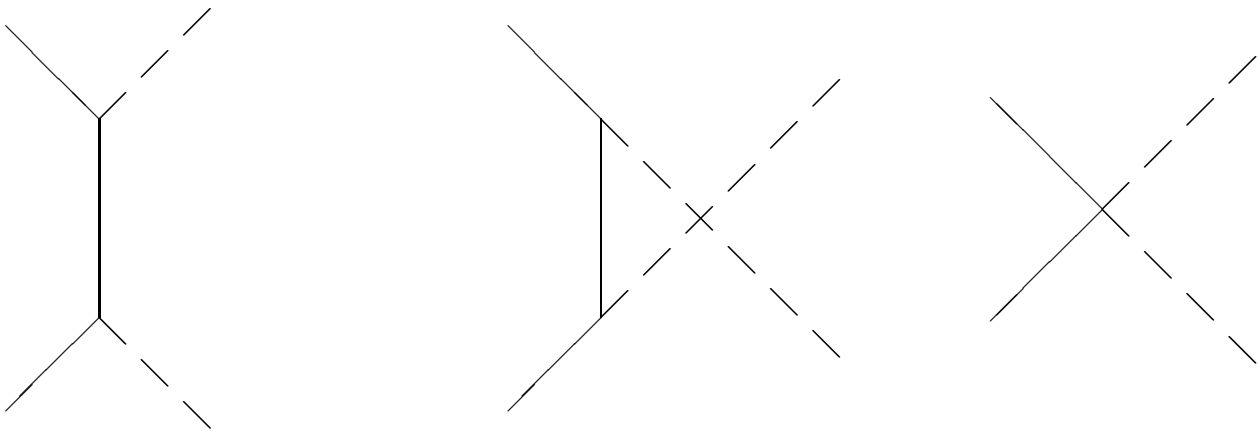
Figure 1: The tree-level graviton Compton diagrams;
solid lines denote massive, spin-$s$ particles, dashed lines denote
gravitons.
\vskip .1in
The corresponding scattering amplitude
involves the propagator $\Pi$ of the massive field $\phi$.
For $s\geq 1$ this propagator contains, in its numerator,
terms proportional to
$1/m^2$, related to the existence of (restricted) gauge invariances in the
$m\rightarrow 0$ limit. These
mass singularities could, in principle, give rise to a scattering amplitude
containing terms ${\cal O}(\mbox{s}^2/m^2M_{Pl}^2)$. Such a scattering
amplitude
would become large, and eventually exceed the unitarity bounds, at
$\sqrt{\mbox{s}} \approx \sqrt{m M_{Pl}}$.
If $m\ll M_{Pl}$ this energy lies far below the Planck scale;
also, the amplitude would have no massless limit due to its dependence on
powers of $m^{-1}$.

The reason why ${\cal O}(\mbox{s}^2/m^2M^2_{Pl})$ terms are absent for
$s = 1 \mbox{,} \; 3/2 \mbox{,} \; 2$ is the following.
The diagrams in fig.~1 giving rise to the dangerous
${\cal O}(\mbox{s}^2/m^2M_{Pl}^2)$
terms have the form\footnotemark~ $J\Pi J$.
\footnotetext{ The `seagull'' diagram in fig.~1 does not contribute to the
leading zero-mass singularity.}
The tensor current $J$ is
obtained by varying the action $S[\phi, e^a_{\mu}]$ with respect to the field
$\phi$, and keeping only terms linear in the fluctuation of the metric or
vierbein\footnotemark~  about
the flat-space background. This vertex is thus universally defined by
\footnotetext{By working in the symmetric gauge $e_{a\mu}=e_{\mu a}$
for the vierbein, we have $e_{\mu a}=\eta_{\mu a} + 1/2 h_{\mu a}$.}
\beq
\left.{\delta S[\phi, e^a_\mu]\over \delta \phi}\right|_L
\equiv J
\label{3}
\eeq
As noticed above, terms proportional to $1/m^2$ in the propagator $\Pi$ are
related to gauge invariance in the massless limit. More precisely
$\Pi\stackrel{\,\,\,o}{\phi}=m^{-2}\stackrel{\,\,\,o}{\phi}$ iff
$\stackrel{\,\,\,o}{\phi}$ is a pure gauge. The
standard form of $\stackrel{\,\,\,o}{\phi}$ for
$s = 1 \mbox{,} \; 3/2 \mbox{,} \; 2$ reads
(throughout, we consider Majorana spinors and real tensors, for simplicity):
\beq
{ s}=1:\;\; \stackrel{\,\,\,o}{\phi}_\mu=\partial_\mu \epsilon
\; \mbox{,} \quad
{ s}=3/2:\;\; \stackrel{\,\,\,o}{\phi}_\mu=\partial_\mu \epsilon
\; \mbox{,} \quad
{ s}=2:\;\; \stackrel{\,\,\,o}{\phi}_{\mu\nu}=\partial_\mu \epsilon_\nu
+\partial_\nu\epsilon_\mu \; \mbox{.}
\label{4}
\eeq
The gauge parameter $\epsilon$ is a  scalar for $s = 1$, a
spinor for $s = 3/2$, and a  vector for $s = 2$.
If the projection of the current $J$ on the vectors
$\stackrel{\,\,\,o}{\phi}$,
denoted by $J \; \cdot \stackrel{\,\,\,o}{\phi}$, has the
form $mX$, with $X$ any operator possessing a smooth $m\rightarrow 0$ limit,
then, by dimensions, no ${\cal O}(\mbox{s}^2/m^2M^2_{Pl})$ terms
will arise in
the scattering amplitude of fig.~1. Here the key observation is that, up to
${\cal O}(h^2)$ terms, $J \; \cdot \stackrel{\,\,\,o}{\phi}$ equals
$\stackrel{\,\,\,o}
{\phi} \cdot \; \delta S[\phi,e^a_{\mu}]/\delta\phi$, due to eq.~(\ref{3}):
we find the projection $J \; \cdot \stackrel{\,\,\,o}
{\phi}$  by varying the action
$S[\phi,e^a_{\mu}]$ under a gauge transformation\footnotemark,
and linearizing in the gravitational field $h_{\mu \nu}$.
\footnotetext{ To be exact, under a restricted gauge transformation of the
massless free Lagrangian,
$S_0[\phi]=\lim_{m\rightarrow 0}S[\phi]$.}
For generic spin this variation contains terms of the form $mX$, hereafter
called ``soft,'' as well as hard terms.
The hard terms do not vanish in the $m\rightarrow 0$ limit\footnotemark.
\footnotetext{The non-vanishing of $J \; \cdot \stackrel{\,\,\,o}{\phi}$ in
the massless limit implies, by~(\ref{3}), that the equations of motion of
a massless high-spin particle are generically inconsistent~\cite{AD1}.}
For
$s =3/2$ and $2$, though,
the hard terms are proportional to the (linearized)
Ricci tensor $R_{\mu\nu}$~\cite{DZ,AD2}; they are absent altogether for $s=1$.
These hard contributions vanish when we impose the free-graviton equations of
motion (that is the linearized Einstein equations in the vacuum) that the
external gravitons in fig.~1 obey.

Having recalled how the lower spins escape the tree-unitarity problem, we will
turn, in the rest of this paper, to the general case. In section 2, we review
the first system, $s=5/2$, in which the full extent of the high-energy-limit
problem is
manifested, and isolate the specific forward scattering amplitude to be
analysed. We will then consider the general integer spin case in section 3, and
find the required non-minimal additions that remove the hard contributions to
that amplitude. The construction is extended to all half-integer spin systems
in section 4, while section 5 contains some concluding remarks and interesting
open problems.
\section{Review of Spin 5/2}
\setcounter{equation}{0}
Before going on to the general case, it is instructive to summarize,
following~\cite{P}, the situation for spin 5/2, where one is first faced with
hard terms involving the full curvature. The system is described,
as in~\cite{SH,B}, by a
symmetric tensor-spinor $\psi_{\mu\nu}$, and an auxiliary spinor
$\chi$. The corresponding action was written, as in~\cite{B,PvN}:
\bea
S&=& \int d^4x  \sqrt{g}\,
\left[ -{1\over 2} \bar{\psi}_{ab}\Dirac\psi_{ab} -\bar{\psi}_{ab}
\gamma_b \Dirac \gamma_c\psi_{ca} + 2\bar{\psi}_{ab}\gamma_bD_c\psi_{ca}
 +{1\over 4} \bar{\psi}_{aa}\Dirac \psi_{bb}
-\bar{\psi}_{aa}D_b\gamma_c\psi_{bc} \; + \right.
\nonumber \\
& & \left. + \; {m\over 2} \; \left( \bar{\psi}_{ab}\psi_{ab}
 -{3\over 4} \bar{\psi}_{ab}\gamma_b \gamma_c\psi_{ca} -{7\over 4}
\bar{\psi}_{aa}\psi_{bb} -{16\over 3} \bar{\chi}\psi_{aa} -{32\over
9}\bar{\chi}\chi \right) \right]\; \mbox{.}
\label{5}
\eea
The conventions on the metric and gamma matrices follow~\cite{PvN}
and the covariant derivative of the field $\psi_{\mu\nu}$ is
\beq
D_\mu\psi_{\nu\rho}=\partial_\mu\psi_{\nu\rho} + {1\over 2}
\sigma_{ab}\omega_{\mu}^{ab}(e)\psi_{\nu\rho} -
\Gamma^\lambda_{\mu\nu}\psi_{\lambda\rho}
-\Gamma^\lambda_{\mu\rho}\psi_{\nu\lambda} \; \mbox{.}
\label{6}
\eeq
The free spin-5/2 Lagrangian possesses a restricted gauge invariance
at $m=0$~\cite{B,S}.
The gauge parameter is a gamma-traceless vector spinor, and the
gauge transformation reads
\beq
\delta\psi_{\mu\nu}=\partial_\mu\epsilon_\nu + \partial_\nu\epsilon_\mu \;
\mbox{,} \;\;\;
\gamma^\mu\epsilon_\mu=0 \; \mbox{.}
\label{8}
\eeq
The free equations of motion of $\psi_{\mu\nu}$ and $\chi$ give, as expected
\beq
\dirac\psi_{\mu\nu}=m\psi_{\mu\nu}\; \mbox{,} \;\;
\gamma^\mu\psi_{\mu\nu}=\partial^\mu\psi_{\mu\nu}=0\; \mbox{,} \;\; \chi=0
\; \mbox{.}
\label{9}
\eeq
In order to see whether ${\cal O}(\mbox{s}^2/m^2M_{Pl}^2)$ terms are present
in the scattering diagrams of fig.~1 one must perform a variation of the
action~(\ref{5})
under the transformation~(\ref{8}), linearize in the gravitational field, and
put $\psi_{\mu\nu}$, $\chi$ and $h_{\mu\nu}$ on shell. A short calculation
gives~\cite{P}
\beq
\delta
S=-\int d^4x 2 \sqrt{g} \,
\bar{\epsilon}_\nu\gamma_\rho\psi_{\lambda\sigma}R^{\nu\lambda\rho\sigma}
\; + \; \mbox{soft terms} \; + \; {\cal O}(h^2) \; \mbox{.}
\label{10}
\eeq
The hard term in this equation is proportional to the Riemann
tensor; thus, it does not vanish on shell, so the scattering amplitude of
fig.~1 does contain ${\cal O}(\mbox{s}^2/m^2M_{Pl}^2)$ terms.

Thus, a minimally coupled light ($m\ll M_{Pl}$) spin-5/2 field
interacts strongly with gravity even at relatively low energies
($\sqrt{\mbox{s}}\approx \sqrt{mM_{Pl}}\ll M_{Pl}$), and the presence of
inverse
powers of the mass also bars us from taking the massless limit.
This scenario is rather bizarre: it would seem more
natural that gravitational interactions remain weak up to Planckian
energies $\sqrt{\mbox{s}}\approx M_{Pl}$, irrespective of any particle's mass.
In~\cite{P} it was shown that in order to implement
this requirement one must add, to the minimal Lagrangian~(\ref{5}),
a non-minimal term
proportional to the Riemann curvature, whose gauge variation cancels
the hard term given in~(\ref{10}).
In other words, in order to obtain gravitational
tree-level amplitudes that may possess a smooth $m\rightarrow 0$ limit
one must add a non-minimal term such that the high-spin current
in~(\ref{3}), i.e.
$J_{\mu\nu}(x)\equiv \delta S[\psi]/\delta \psi_{\mu\nu}(x)|_L$,
is conserved up to terms proportional to $m$.

The required non-minimal term turned out to be\footnotemark~ \cite{P}
\beq
-\frac{1}{2}m
\int d^4x  \sqrt{g}\,  \bar{\psi}_{\mu\nu}R^{\pm\,\mu\rho\nu\sigma}
\psi_{\rho\sigma}, \;\;\;\;
R^{\pm\,\mu\rho\nu\sigma}=R^{\mu\rho\nu\sigma} \pm {1\over 2}
\gamma^5\epsilon^{\nu\sigma\alpha\beta}R^{\mu\rho}_{\alpha\beta}.
\label{11}
\eeq
\footnotetext{At $R_{\mu\nu}=0$ the quantities $R^{\pm\,\mu\rho\nu\sigma}$
reduce, on
fermionic states of definite chirality, to the self-dual (or anti- self-dual)
components of the Weyl tensor $C^{\mu\rho\nu\sigma}$.}

Actually the improvement~(\ref{11}) does not by itself accomplish the desired
goal of removing all singular terms from the scattering amplitude. The
fermionic current $J_{\mu\nu}$ has two parts: the pure spin 5/2, transverse and
$\gamma$-traceless one, and its longitudinal complement, the (lower spin)
part in the
direction of the gauge variation~(\ref{8}). The non-minimal addition has
ensured that the contribution of the latter is well-behaved in the massless
limit. However~(\ref{11}) also contributes a (singular) part to the transverse
current, and the latter will contribute already against the non-singular parts
of the propagator $\Pi$.
We believe that in order to
cancel this mass singularity one would be forced -- at least --
to add other massive
fields, of spin larger than 5/2. Probably the process never stops until an
entire Regge trajectory has been added. A similar conjecture has been made
in~\cite{CLT},
in the case of purely gravitational interactions of elementary particles,
and has been used in~\cite{SS}
as one of the early motivations to consider
string theory as the natural setting of a perturbative theory of quantum
gravity.
In terms of a field equation analysis, the addition~(\ref{11}) only
ensures that
the longitudinal projection of the matter equations obeys Bianchi identities
devoid of explicit curvature terms; but it does not remove such (singular)
terms
in the transverse complement.

The implementation of tree-level unitarity on all scattering amplitudes
is a most interesting
program, especially because the (conjectured)
need for new (infinitely many) degrees of
freedom seems to lead naturally to string theory. On the other hand this
program seems difficult to achieve concretely at present,
and thus we prefer to
adopt another strategy, namely to find scattering amplitudes that do not
depend on the transverse current, and implement tree-level unitarity only on
those amplitudes.
In other words, we prefer to divide the implementation of tree-level
unitarity into two parts. One will give rise to constraints on the
interactions of individual
high-spin particles alone; the other will not only further constrain
these interactions but also require the introduction of
new degrees of freedom.

Fortunately, an amplitude independent of the transverse part of the current
$J$ is known. It can be written for particles of arbitrary spin,
and reads
\beq
f_-(E,{ s}_3) = {1\over 2E} \; \left[
     f \left( E,{ s}_3,+ \right) - f \left( E,{ s}_3,- \right) \right]
\; \mbox{.} \label{11'}
\eeq
It is the difference between the positive- and negative-helicity
forward elastic scattering amplitude of a graviton of energy $E$ off
a spin-$s$ particle at rest; $-s\leq { s}_3\leq s$ is the helicity of the
spin-$s$ particle. The equivalent amplitude, in the case of photon
scattering off a spin-$s$ target,  was used by Weinberg~\cite{W} to constrain
the form of electromagnetic interactions of high-spin particles, by using
appropriate dispersion relations.
The natural question is then whether
the cancellation of hard terms in~(\ref{11'})
is specific to spin 5/2 or whether it holds in general.
We shall show how the results of~\cite{P} can be extended
to massive particles of {\em any}
spin, provided suitable gravitational quadrupoles
are added to the Lagrangian: in this respect, spin 5/2 is not special.

The fact that the forward elastic scattering amplitude $f_-(E,s_3)$
is tree-level-unitary means that there exist a
way of defining the gravitationally-coupled action such that the current $J$,
defined in eq.~(\ref{3}) is softly broken. This means, as explained before,
that its divergence is proportional to $m$. If one could take the
$m\rightarrow 0$ limit of~(\ref{3}), one would then obtain an exactly
conserved tensor current, which would imply the existence of a large
``softly broken'' higher-spin symmetry. To justify this conjecture one may
appeal to the case of the standard-model W-bosons in an external e.m.
field: in that case the role of the currents $J$ is played by
the off-diagonal components of the $SU(2)$ gauge group\footnotemark.
\footnotetext{ The link between tree-level
unitarity and gauge symmetry indeed holds for any system of interacting
particles of spin $\leq$ 1~\cite{CLT,LS}.}
The $m\rightarrow 0$ limit may be legitimate in a completely consistent theory
of interacting higher spins, as string theory is conjectured to be. The
massless limit there may correspond to the tensionless-string limit
$\alpha'\rightarrow \infty$~\cite{Gr,ILS} or to a new topological phase
of gravity~\cite{Wi}.
\section{Integer Spins}
\setcounter{equation}{0}
To compute the amplitude~(\ref{11'}) for arbitrary spin we need an
explicit form of the free massive, spin-$s$ field Lagrangian; we use
that of~\cite{SH}, which is
written in terms of symmetric, traceless
tensors $\phi^{p}$ of rank $p = s \mbox{,} \; s-2 \mbox{,} \; s-3 \mbox{,}
\; \ldots \mbox{,} \; 0$ :
\bea
L & = & \frac{1}{2}  \phi^{s} \left( \partial^{2} - m^{2} \right)
          \phi^{s}
     +  \frac{s}{2}  \phi^{s} \cdot \stackrel{\leftarrow}{
                    \partial} \partial \cdot \phi^{s}  +
      \left.\frac{s \left( s-1 \right)^{2} }{2 s - 1}
\right[  \phi^{s-2} \partial \cdot \partial \cdot \phi^{s}  +
 \nonumber \\
       &  & -\left. \frac{1}{2}  \phi^{s-2} \left( \partial^{2} -
            \frac{s}{s-1}  m^{2} \right) \phi^{s-2}
       + \frac{ \left( s-2 \right)^{2} }{2 \left( 2 s - 1
                    \right)}
           \phi^{s-2} \cdot \stackrel{\leftarrow}{ \partial}
               \partial \cdot \phi^{s-2}  +  \ldots  \right],
\label{12}
\eea
where $\partial \cdot \phi^{p} \equiv \partial^{\mu} \phi^{p}_{\mu
\mu_{2} \ldots \mu_{p}}$ .
This Lagrangian involves the minimal number of auxiliary (non-propagating)
fields needed in order to describe the propagation of a pure spin-$s$ field.
Since we shall
consider only perturbative $S$-matrix elements, any other consistent Lagrangian
should give rise to equivalent results.
The equations of motion obtained
from~(\ref{12}) are:
\beq
\phi^{p<s} = 0, \; \;\;
\left( \partial^{2} - m^{2} \right) \phi^{s} = 0
\; \mbox{,} \; \;\; \partial \cdot \phi^{s}  = 0 \; \mbox{.}
\label{13}
\eeq
The Lagrangian~(\ref{12}) may now be coupled to gravity using the
substitution
\beq
\partial_{\mu}\rightarrow D_{\mu}\equiv \partial_\mu + \omega_\mu^{ab}L^{ab},
\;\;\; \eta^{\mu\nu}\rightarrow
g^{\mu\nu}=e^{\mu}_a e^{\nu\, a} \; \mbox{.}
\label{1}
\eeq
The resulting minimal (with respect to the derivative ordering in~(\ref{12}))
action\footnotemark
\footnotetext{
Minimal coupling to any gauge theory, including gravity, is of course only
defined uniquely for systems of first derivative order (fermions or first-order
form of bosons); for bosons of non-zero spin, different order of derivatives
will differ by curvature terms. This ambiguity is unimportant -- the real
question is whether any given ordering, corrected if necessary by non-minimal,
``Pauli,'' terms proportional to the curvature or its derivatives, can lead to
the desired amplitude behavior.
}
 $S_m[\phi, g_{\mu\nu}]$ gives
rise to a current $J$ as in~(\ref{3}).
As we explained in the Introduction (and illustrated for the case of
spin 5/2) the longitudinal part of $J$ is associated with a
restricted gauge invariance of the massless free Lagrangian.
More precisely, one varies $S_m[\phi,g_{\mu\nu}]$ with respect to the gauge
transformation~\cite{FPT}:
\bea
\delta \phi^{s}_{\mu_{1} \ldots \mu_{s}} & = &
\partial_{\left( \mu_{1} \right.}
\epsilon_{\left. \mu_{2} \ldots \mu_{s} \right)} \; - \;
\frac{s - 1}{2 s} \; g_{\left( \mu_{1} \mu_{2} \right.} \partial^{\lambda}
\epsilon_{\left. \mu_{3} \ldots \mu_{s} \right) \lambda}
\nonumber \\
\delta \phi^{s-2}_{\mu_{1} \ldots \mu_{s-2}} & = &
\frac{2 s - 1}{s^{2}} \;
\partial^{\lambda} \epsilon_{\mu_{1} \ldots \mu_{s-2} \lambda}
\nonumber \\
\delta \phi^{p}_{\mu_{1} \ldots \mu_{p}} & = &
0 \quad \quad \mbox{if} \quad p < s - 2 \; \mbox{,}
\label{14}
\eea
where the parentheses mean total (normalized) symmetrization and the gauge
parameter $\epsilon_{\mu_{1} \ldots \mu_{s-1}}$ is a  symmetric, traceless,
rank-$s-1$ tensor, obeying the equation
\beq
\partial^{\mu} \partial^{\lambda} \epsilon_{\mu
\lambda \mu_{3} \ldots \mu_{s-1}} \; = \; 0 \; \mbox{.}
\label{15}
\eeq
The variation of the action under the transformation~(\ref{14})
gives, after the free equations of motion for the fields $\phi^{p}$
and $h$ are used,
\beq
\delta S_m = \int d^4x \sqrt{- g}
2 \left( s - 1 \right) \epsilon_{\gamma}^{\;\; \mu_{2} \ldots
                    \mu_{s-1}} R^{\alpha \beta \gamma \delta}
 \partial_\alpha \phi^{s}_{\beta \mu_{2} \ldots \mu_{s-1} \delta}
        +
 \mbox{soft terms}  + {\cal O}(h^2).
\label{16}
\eeq
The hard terms, proportional to the Riemann tensor and to
$\left( s - 1 \right)$, vanish for spin $s = 1$ but
do not vanish on shell. Equation~(\ref{16}) is simply the projection of
the current $J$ over the gauge directions: it is
proportional to the longitudinal part of $J$.
Since the propagator $\Pi$ of the massive field $\phi$ is singular as $1/m^2$
along those gauge directions, to obtain a well-behaved scattering
amplitude $f_-(E,s_3)$ one must add to the minimal Lagrangian a new term
$S_{nm}$ such that $\delta S_m + \delta S_{nm}$ is soft. This term is
\beq
S_{nm} =  \frac{s \left( s - 1 \right)}{2}  \int d^4x  \sqrt{- g}
        \phi_{\alpha \gamma}^{s \; \; \; \;\mu_{3} \ldots \mu_{s}}
       R^{\alpha \beta \gamma \delta}
\phi^{s}_{\beta \delta \mu_{3} \ldots \mu_{s}}.
\label{17}
\eeq
The variation of the new non-minimal action $S=S_m +S_{nm}$ gives
\bea
\delta S & = & - \int d^4x  \sqrt{- g}
            \left( s - 1 \right) \left( s - 2 \right)
          \epsilon_{\alpha \gamma}^{ \;\;\;\;\mu_{3} \ldots \mu_{s-1}}
         \left( \partial^{\lambda} R^{\alpha \beta \gamma \delta} \right)
   \phi^{s}_{\beta \delta \mu_{3} \ldots \mu_{s-1} \lambda}
        + \nonumber \\
   &  & + \mbox{soft terms} + {\cal O}(h^2)  \mbox{.}
\label{18}
\eea
Again, one finds that there are hard terms non-vanishing on shell,
but now they are zero for spin $s < 3$. It is interesting to
notice that this result, in
agreement with~\cite{ALR}, can be obtained directly from some ``minimal''
action by exploiting the ambiguity of the minimal coupling,
i.e. without introducing the term~(\ref{17}).
In fact,
if we minimally couple the free Lagrangian~(\ref{12}) to gravity after
integrating the second term by parts, and exchanging the order of
the derivatives, namely if we use the term
$\frac{s}{2}
\partial_\beta\phi^{s \; \alpha \mu_{2} \ldots \mu_{s}}
\partial_{\alpha} \phi^{s \; \beta}_{\; \mu_{2} \ldots \mu_{s}}$
instead of
$\frac{s}{2}
\partial_\alpha\phi^{s \; \alpha \mu_{2} \ldots \mu_{s}}
\partial_{\beta} \phi^{s \; \beta}_{\;\;\;\; \mu_{2} \ldots \mu_{s}}$
,
the variation~(\ref{18}) is obtained directly.
By using the identity
\beq
D_{\alpha} D_{\beta} \; \phi^{s}_{\mu_{1} \ldots \mu_{s}} \equiv
\left( D_{\beta} D_{\alpha} + \left[ D_{\alpha} \mbox{,}
D_{\beta} \right] \right) \phi^{s}_{\mu_{1} \ldots \mu_{s}} \equiv
D_{\beta} D_{\alpha} \; \phi^{s}_{\mu_{1} \ldots \mu_{s}} + \;
s \; R_{\alpha \beta \mu_{1}}^{\quad \; \; \tau} \;
\phi^{s}_{\tau \mu_{2} \ldots \mu_{s}}
\label{18'}
\eeq
one can verify that the minimal action defined in this way
is different from the previous non-minimal one $S_m + S_{nm}$
only through terms proportional to the Ricci tensor, that is through terms
vanishing on graviton shell.
This confirms what was sketched in the Introduction:
the minimal coupling gives rise to hard terms in the
scattering diagrams of fig.~1 only for spins $s \geq 3$ .
Moreover, this new minimal action is equivalent to the first-order
formalism~\cite{AD2};
in fact, its variation under the transformation~(\ref{14}), with
standard derivatives replaced by covariant ones and
by using the identity (our convention is $R^\tau_{\mu\alpha\beta}\sim
+\partial_\alpha \Gamma^\tau_{\beta\mu}$ and the signature is mostly minus)
$D^{\alpha} R_{\alpha \beta \gamma \delta} \equiv
D_{\delta} R_{\gamma \beta} -
D_{\gamma} R_{\beta \delta} $, gives
\bea
\delta S & = &\int d^4 x\sqrt{-g}\left\{
- \left( s - 1 \right)  \left( s - 2 \right) \;
       \epsilon_{\alpha \gamma}^{\; \; \; \; \mu_{4} \ldots \mu_{s}}
          \; \left[
      \left( D^{\lambda} R^{\alpha \beta \gamma \delta} \right)
          \phi^{s}_{\lambda \beta \delta \mu_{4} \ldots \mu_{s}} \; +
       \; 2 \; R^{\alpha \beta \gamma \delta} D^{\lambda}
    \phi^{s}_{\lambda \beta \delta \mu_{4} \ldots \mu_{s}}
        \right] \right.\; + \nonumber \\
 & & + \left( s - 1 \right)
     \; \epsilon_{\alpha }^{\; \; \mu_{3} \ldots \mu_{s}}  \; \left[
    - 2 \; \left( D^{\lambda} R^{\alpha \beta}  \right) \;
           + \; \left( D^{\alpha} R^{\lambda \beta}  \right)
   \; - 2 \; R^{\alpha \beta} D^{\lambda}
     \right] \phi^{s}_{\lambda \beta \mu_{3} \ldots \mu_{s}} + \nonumber \\
    & & \left.+ \; m^{2} \; \epsilon^{\mu_{2} \ldots \mu_{s}}
         D^{\lambda} \phi^{s}_{\lambda \mu_{2} \ldots \mu_{s}}\right\} ,
\label{18''}
\eea
which confirms and extends the result found in~\cite{AD2}.

Now, to eliminate the unwanted curvature derivative
terms in~(\ref{18}), one needs to
add truly non-minimal terms to the action (either the minimal one obtained
with the order of derivatives specified above or the non-minimal one
previously introduced, that is $S =  S_m + S_{nm}$). After a straightforward
but tedious calculation they turn out to be the sum $\Delta S= \Delta S_1 +
\Delta S_2 +
\Delta S_3$ of three contributions
\bea
\Delta S_1 & = &
\frac{s \left( s - 1 \right) \left( s - 2 \right)}{2
m^{2}}
\int d^4x \sqrt{-g}  \left[ \partial^\lambda
\phi_{\alpha \gamma \lambda}^{s \;
		\; \; \; \mu_{4} \ldots \mu_{s}}
             \left(\partial^{\rho}
        R^{\alpha \beta \gamma \delta}\right)
     \phi_{\beta \delta \rho \mu_{4} \ldots
                \mu_{s}}^{s}\right. \nonumber \\ & & \left.
- \; \phi_{\alpha \gamma \lambda}^{s \; \; \; \;
                 \mu_{4} \ldots \mu_{s}}
     R^{\alpha \beta \gamma \delta}  \partial^{\lambda} \partial^{\rho}
\phi^{s}_{\beta \delta \rho \mu_{4} \ldots \mu_{s}}\right],
\nonumber \\
\Delta S_2 & = &
\frac{2s \left( s - 1 \right)^2 \left( s - 2 \right)}{
m^{2}\left( 2s-1\right)}
\int d^4x \sqrt{-g}  \left[ \phi_{\alpha}^{s-2 \; \; \mu_{2}
                            \ldots \mu_{s-2}} \left(
            \partial^{\rho} R^{\alpha \beta \gamma \delta} \right)
        \partial_{\gamma} \phi^{s}_{\rho \mu_{2}
                 \ldots \mu_{s-2} \beta \delta} \right]
\; \mbox{,} \nonumber \\
\Delta S_3 & = &
\frac{s \left( s - 1 \right)^2 \left( s - 2 \right)\left(s-3\right)}{
m^{2}\left(2s-1\right)}
\int d^4x \sqrt{-g}
\left[  \phi_{\alpha \gamma}^{s-2 \; \;
                  \mu_{3} \ldots \mu_{s-2}}
          \left( \partial^{\lambda} \partial^{\rho}
            R^{\alpha \beta \gamma \delta} \right)
         \phi^{s}_{\beta \delta \mu_{3} \ldots \mu_{s-2}
                     \lambda \rho}
             \right] .
\nonumber \\ & & \label{19}
\eea
In this way we find the variation of the complete action $S_T=S+\Delta S$ to be
\bea
\delta S_T & = & \int d^4x \sqrt{-g}
 \frac{\left( s - 1 \right) \left( s - 2 \right)}{m^2} \;
        \epsilon_{\alpha \gamma}^{\; \; \; \; \mu_{3} \ldots \mu_{s-1}}
              \left( \partial^{2} - m^{2} \right)
           \left[ \left( \partial^\rho R^{\alpha \beta \gamma \delta} \right)
       \phi^{s}_{\beta \delta \mu_{3} \ldots \mu_{s-1} \rho}
                  \right] + \nonumber \\
& & +\mbox{soft terms} \; + \; {\cal O}(h^2) \; \mbox{.} \label{21}
\eea
In this equation there is only one hard term (vanishing for spin
$s < 3$) and it is proportional to
$\partial^{2} - m^{2}$. When inserted in the expression for the scattering
amplitude $J\Pi J$, it gives rise to a local ``seagull'' term that
can be cancelled by adding {\em local} counterterms
${\cal O}(h^2)$, and therefore it can be neglected. This is the new feature
that appears at spin 3: the longitudinal part of the current $J$
need not be soft, but it may contain hard parts as long as they are
proportional to the equations of motion of the field $\phi$; these only
give rise to ``seagull'' diagrams that can
be consistently ignored. Another way of stating this property is that
terms proportional to the equations of motion of $\phi$ can be eliminated by
a {\em local} field redefinition of $\phi$, which does not change the form
of the free Lagrangian.
Finally, notice that the $\Delta S$ of~(\ref{19}) vanish on the free
$\phi$-shell, where $\partial \cdot \phi^s=0=\phi^{s-2}$.
This means that they do not
contribute to such physical
quantities as the higher gravitational multipoles.
\section{Half-Integer Spins}
\setcounter{equation}{0}
To study the fermionic systems, we follow the formulation of~\cite{SH}
(slightly different from that of section 2) and write the
free flat-space Lagrangian for massive particles of half-integer spin
$s = n + \frac{1}{2}$ as
\bea
L & = & \bar{\psi}^{n} \left( \dirac - m \right) \psi^{n}  +
      \left. \frac{2 n^{2}}{2 n + 1} \right[
           \bar{\psi}^{n-1} \partial \cdot  \psi^{n}  -
            \bar{\psi}^{n} \cdot \partial \psi^{n-1} +
            \nonumber \\
   & & +\left.  \bar{\psi}^{n-1}
        \left( \dirac + \frac{n + 1}{n} m \right) \psi^{n-1} +
      \ldots  \right] .
 \label{22}
\eea
The fields $\psi^{p}$ are the minimal set needed to
represent the physical degrees of freedom of a pure spin-$s$
free massive particle. They are symmetric $\gamma$-traceless
tensor-spinors of rank $p = n \mbox{,} \; n-1 \mbox{,} \; n-2
\mbox{,} \; \ldots \mbox{,} \; 0$ ; for rank lower than $n-1$,
they appear in $L$ with multiplicity $2$ (namely $\psi^{i \mbox{,}
\; p}$ and $i = 1 \mbox{,} \; 2$ ). Here the metric is the
Pauli metric $g^{\mu \nu} =  \delta^{\mu \nu}$ and the
convention on the gamma matrices (see~\cite{PvN})
is:
\beq
\gamma^{\mu} \gamma^{\nu} + \gamma^{\nu} \gamma^{\mu} =
           2 \delta^{\mu \nu} \; \mbox{,} \quad
\gamma^{\mu^{\dagger}} = \gamma^{\mu}  \; \mbox{,} \quad
\gamma^{5} = \gamma^{1} \gamma^{2} \gamma^{3} \gamma^{4} \; \mbox{.}
\label{23}
\eeq
Note that since $\psi^p$ is symmetric and $\gamma$-traceless, it is
automatically also traceless:
$\psi^{p\;\mu}_{\mu\mu_{3} \ldots \mu_{p}} = 0$.
The free equations of motion derived from~(\ref{22}) are:
\beq
\left( \dirac - m \right) \psi^{n}  =  0;\;\;
\partial \cdot  \psi^{n} =  0,\;\;
\psi^{n-1} =0= \psi^{i \mbox{,} \; p},\;\; p<n-1,\;i=1,2.
\label{23a}
\eeq
The free massless Lagrangian possesses a restricted gauge
transformation~\cite{FPT} given by:
\bea
\delta\psi^{n}_{\mu_{1} \ldots \mu_{n}} & = &
            \frac{2 n \left( n + 1 \right)}{2 n + 1}
       \; \partial_{\left( \mu_{1} \right.} \epsilon_{\left. \mu_{2}
                       \mu_{3} \ldots \mu_{n} \right)} -
    \frac{n}{2 n + 1} \; \gamma_{\left( \mu_{1} \right.}
              \dirac \epsilon_{\left. \mu_{2}
                  \mu_{3} \ldots \mu_{n} \right)} \nonumber \\
\delta\psi^{n-1}_{\mu_{1} \ldots \mu_{n-1}} & = &
          \dirac \; \epsilon_{\mu_{1} \ldots \mu_{n-1}} \nonumber \\
\delta\psi^{i \mbox{,} \; p} & = & 0 \quad \quad i = 1 \mbox{,} \; 2
\; \; \mbox{and} \; \; \; \forall \; p < n - 1
         \label{25}
\eea
where $\epsilon$ is a rank $n-1$ symmetric tensor-spinor that is
$\gamma$-traceless and divergenceless.
While half-integer spins have unique minimal coupling to gravity, they share
the tree-unitarity problems of spin 5/2 discussed in Section 2.
In fact, the variation of the minimal action $S_m$
under the transformation~(\ref{25}),
linearized in the gravitational field and with the fields $\psi^{p}$
and $h$ put on free shell, gives
\bea
\delta S_m & = & \int d^4x  \sqrt{g}\,
             \frac{4 n \left( n^2 - 1 \right)}{2 n + 1}
         \bar{\epsilon}^{\alpha \mu_{3} \ldots \mu_{n}}
           \gamma^{\lambda}  R_{\alpha \beta \lambda \delta}
      \psi^{n\; \beta \delta}_{ \quad \; \mu_{3} \ldots \mu_{n}}
           + \mbox{soft terms} + {\cal O}(h^2)
\nonumber \\
   & = & \int d^4x  \sqrt{g}\,
\frac{4 n \left( n^2 - 1 \right)}{2 n + 1}
           \frac{1}{m}   \bar{\epsilon}^{\alpha \mu_{3} \ldots
                  \mu_{n}}
            R^{\; +}_{\alpha \beta \lambda \delta}
           \partial^{\lambda} \psi^{n \; \beta \delta}_{ \quad
                     \; \mu_{3} \ldots \mu_{n}}  + \nonumber \\
           & & +\mbox{soft terms}  +  {\cal O}(h^2).
\label{27}
\eea
By performing the same procedure as used in the previous section, we
can remove the hard term in eq.~(\ref{27}); here we add
the non-minimal term
\beq
S_{mn}=
       \frac{n(n-1)}{2m} \int d^4x\sqrt{g}\,
\bar{\psi}^{n \;  \alpha \gamma \mu_{3} \ldots \mu_{n}} \;
         R^{\; +}_{\alpha \beta \gamma \delta} \;
    \psi^{n \; \beta \delta}_{ \quad \; \mu_{3} \ldots \mu_{n}}.
\label{29}
\eeq
This result confirms~\cite{P}; in analogy with the bosonic case,
for spin $s \geq 7/2$ (i.e. $n \geq 3$), the variation of the new
non-minimal action gives rise to additional hard terms.
They read
\beq
\delta S= - \frac{2 n \left( n^{2} - 1 \right) \left( n - 2 \right)}{
      m\left( 2 n + 1 \right) }
\int d^4x \sqrt{g}\,
    \bar{\epsilon}^{\alpha \gamma \mu_{3} \ldots \mu_{n-1}}
   \left( \partial^{\lambda} R^{\; +}_{\alpha \beta \gamma \delta} \right)
    \psi^{n \; \beta \delta}_{ \; \quad \mu_{3} \ldots \mu_{n-1}
          \lambda} \mbox{.}
\label{30}
\eeq
These terms can be cancelled, as in the bosonic case, by introducing extra
terms, {\em which again vanish on shell~(\ref{23a})}:
\beq
\Delta S =
\frac{2 n \left( n^{2} - 1 \right) \left( n - 2 \right)}{
m^3 \left( 2 n + 1 \right) }
\int d^4x \sqrt{g}\,
\left\{ \bar{\psi}^{n-1 \;  \alpha \gamma \mu_{3} \ldots \mu_{n-1}}
          \left[ \partial^{\lambda}\left(m+\dirac\right)
R^{\; -}_{\alpha \beta \gamma \delta}
               \right]
  \psi^{n \; \beta \delta}_{ \quad \; \mu_{3} \ldots \mu_{n-1} \lambda}
\right\}.
\label{31}
\eeq
The variation of the total action $S_m + S_{nm} +\Delta S$ now gives
\bea
\delta S & = & \int d^4x \sqrt{g}\,
 \frac{2 n \left( n^{2} - 1 \right) \left( n - 2 \right)}{
\left( 2 n + 1 \right) m^{3}}
    \bar{\epsilon}^{\alpha \gamma \mu_{3} \ldots \mu_{n-1}}
  \left( \partial^{2} - m^{2} \right)  \left[
    \left( \partial^{\lambda} R^{\; +}_{\alpha \beta \gamma \delta}
  \right) \psi^{n \; \beta \delta}_{ \; \; \mu_{3}
         \ldots \mu_{n-1} \lambda} \right] +
\nonumber \\
& & +\mbox{soft terms}  +  {\cal O}(h^2)  \mbox{.}
\label{32}
\eea
The remarks made about integer spin
at the end of the previous section apply here as well: the
terms proportional to the equations of motion of $\psi$ are irrelevant since
they give rise to local ``seagull'' diagrams, and the additional terms in
the Lagrangian, given in~(\ref{31}), {\em do not} contribute to the
gravitational multipoles of the spin-$s$ particle since they vanish on shell
and can be eliminated
by a local redefinition of $\psi$, which changes neither the kinetic term
nor the $S$-matrix.

\section{Conclusions}
\setcounter{equation}{0}
We have established for all massive, higher-spin fields that one can adequately
soften the behavior of their forward elastic scattering amplitudes
$f_-(E,s_3)$ to maintain
tree-level
unitarity up to the
Planck scale. This was carried out explicitly and uniformly
through addition of non-minimal terms proportional to the Riemann (or Weyl)
tensor and its derivatives. Equivalently, this means that the projection in the
gauge direction (of the massless limit) of the improved matter field equations
obeyed Bianchi identities not involving explicit curvature terms. We noted,
however, that the complementary, gauge-orthogonal, contributions to the
vertices (as well as the propagators)
now contain inverse powers of mass and hence the corresponding
amplitudes will still violate unitarity at lower scales; but conjecture that
improvements here (if they can be achieved at all) will
require taking into account
couplings of infinite towers of massive particles.
We emphasize that the transverse and longitudinal currents give rise to
different pathologies. In particular, since $f_-(E,s_3)$ does {\em not}
depend on spins higher than $s$,
it is only
sensitive to the longitudinal current. Thus, a {\em necessary} condition for
tree-level unitarity is that this current be improved so as to become
softly broken. On the
other hand the transverse-current pathologies could be removed, at least in
principle, by adding new higher-spin massive states, an addition that {\em
cannot} help the longitudinal current. It must be, and was, improved entirely
within the single-field framework.
Two open questions are
particularly interesting. First, how do our new
terms compare
with the way closed string theory organizes its higher-spin excitations,
in the domain in which such an
expansion is valid? Its beta function, expanded to
include these massive modes, should show both the longitudinal compensation
terms given here as well as the more complicated transverse corrections we
conjectured. The second question involves the gravitational quadrupole
interactions implied by our non-minimal terms. Is there a universal value,
similar to $g=2$ in the corresponding electromagnetic higher-spin
analysis, of the total static gravitational quadrupole moments?
\vskip .1in
\noindent
Acknowledgements\\
\noindent
S.D. was supported in part by NSF under grants PHY 93-15811, 92-45316 and by
the  Ambrose Monell Foundation, at the Institute for Advanced Study;
M.P. was supported in part by NSF under grant PHY 93-18781.

\end{document}